\documentclass[opre,nonblindrev]{informs3}  

\DoubleSpacedXI 

\usepackage{tikz}
\usepackage{gensymb}
\usepackage[utf8]{inputenc} 
\usepackage[T1]{fontenc}    

\usepackage{booktabs}       
\usepackage{amsfonts}       
\usepackage{nicefrac}       
\usepackage{bm}
\usepackage{wrapfig}
\usetikzlibrary{fit,positioning}
\usepackage{amsmath}
\usepackage{subcaption}
\usepackage[hyphenbreaks]{breakurl}
\usepackage{hyperref}       

\usepackage{natbib}
\bibpunct[, ]{(}{)}{,}{a}{}{,}%
\def\bibfont{\small}%
\graphicspath{{./figures/}}
\newcommand{\vx}{\mathbf{x}}
\newcommand{\vz}{\mathbf{z}}

\newcommand{\curly}[1]{\left\{#1\right\}}
\renewcommand{\Pr}{\mathbb{P}}

\DeclareMathOperator*{\minimize}{minimize}

\TheoremsNumberedThrough     
\ECRepeatTheorems

\EquationsNumberedThrough    

\TITLE{Detecting Influence Campaigns in Social Networks Using the Ising Model}
\RUNTITLE{Detecting Influence Campaigns in Social Networks Using the Ising Model}

\RUNAUTHOR{Guenon des Mesnards and Zaman}
\ARTICLEAUTHORS{%
	\AUTHOR{Nicolas Guenon des Mesnards}
	\AFF{%
		Operations Research Center, Massachusetts Institute of Technology, 
		Cambridge, MA~  02139, 
		\EMAIL{ngdm@mit.edu}
	}
	\AUTHOR{Tauhid Zaman}
	\AFF{%
		Sloan School of Management, Massachusetts
		Institute of Technology,   
		Cambridge, MA~ 02139,  
		\EMAIL{zlisto@mit.edu}
	}
}

\ABSTRACT{We consider the problem of identifying coordinated influence campaigns conducted by
	automated agents or bots in a social network.  We study several different Twitter datasets which contain such campaigns and find that the bots exhibit heterophily - they interact more with humans than with each other.  We use this observation to develop a probability model for the network structure and bot labels based on the Ising model from statistical physics.  We present a method to find the maximum likelihood assignment of bot labels by solving a minimum cut problem.  Our algorithm allows for the simultaneous detection of multiple bots that are potentially engaging in a coordinated influence campaign, in contrast to other methods that identify bots one at a time.  We find that our algorithm is able to more accurately find bots than existing methods when compared to a human labeled ground truth.  We also look at the content posted by the bots we identify and find that they seem to have a coordinated agenda.
}

\SUBJECTCLASS{  
	Social networks, Influence campaigns, Ising model.
}

\renewcommand{\theARTICLETOP}{}

\begin{document}

	\maketitle
	
	%
	
	
	\section{Introduction}\label{sec:intro}
	Social networks face the challenge posed by automated bots which
	create spam in the network and result in a degraded user experience.  However,
	recently these bots have become a serious threat to democracies.
	There have been multiple reports alleging that foreign actors attempted to penetrate U.S. social networks in order to manipulate elections \citep{ref:russianbots,shane2017fake,guilbeault2016twitter,byrnes2016bot,ferrara2017disinformation}.  The perpetrators used bots to share  politically polarizing content, much of it fake news, in order to amplify it and extend its reach, or directly interacted  with humans to promote their agenda.    While no one knows exactly how many people were impacted by these influence campaigns, it has still become a concern for the U.S. and many other governments \citep{ref:russianbots_govtresponse,ref:russianbots_feinstein}.    
	
	Social network counter-measures are needed to combat these coordinated influence campaigns.  Conventional methods of bot detection may not be sufficient because they can be fooled by modifying certain elements of bot behavior, such as the frequency of posting or sharing content.  However, because many of these bots are coordinated, they may exhibit joint behaviors which are difficult to mask and which allow for more accurate and robust detection.  These behaviors may not be observable by looking at accounts in isolation.  Therefore,   conventional algorithms which focus on individual detection may not find these bots.  What is needed  is an algorithm that can simultaneously detect multiple bots.

	\subsection{Our Contributions}\label{sec:our_contrib}
	 We provide evidence that modern bots in the social network Twitter coordinate their attacks. They do not create original content, but rather  amplify certain human users by disproportionately retweeting them. We find that the bots exhibit \emph{heterophily}, which means they interact with humans much more frequently than they interact with each other \citep{rogers2010diffusion}.  This is the opposite of the more common homophily phenomenon, where similar individuals interact with each other.
	
	We use heterophily to design a new algorithm for bot detection.
	We utilize the Ising model from statistical physics to model the network structure and bot labels.  We show that the maximum likelihood estimation of the bot labels in this Ising model reduces to solving a minimum cut problem, which allows for efficient estimation. Our algorithm scales to large networks and is much faster than modern bot detection algorithms (see Table \ref{table:runtimeVSaccuracy}).
	
	A natural application of our algorithm is understanding the agenda of the bots' influence campaigns.    We do this by analyzing the content of the detected bots.
	We find that the bots seem to promote certain geo-political issues.
	
	We emphasize that our algorithm  is \emph{language agnostic} and relies only on the interaction network.  Therefore, it can be used on social networks in many different countries without having prior knowledge of the language, culture, or individuals involved. Also, our algorithm is not restricted to Twitter, and can be used on any online platform where individuals can interact with one another (Facebook, Reddit, Instagram, etc.).

	\subsection{Previous Work}\label{sec:previous_work}
	 Originating from spam detection \citep{sahami1998bayesian}, a significant subset of bot detection algorithms are based on recognizing patterns of automation \citep{heymann2007fighting, gyongyi2004combating, gyongyi2005web, ntoulas2006detecting, castillo2007know}. It was observed by \cite{malmgren2009characterizing} and \cite{raghavan2014modeling} that human beings present specific circadian patterns which can further be used for strategic timing or promotion campaigns \citep{ li2014detecting, modaresistrategic}, or, in our case, to discriminate humans from bots \citep{zhang2011detecting, chu2012detecting, ghosh2011entropy}. Posting at random times makes bots easily detectable, as shown in \cite{gianvecchio2011humans}, but bots that retweet tweets of actual humans will replicate human rhythms with a lag. Modern bots are no longer unintelligent spam accounts, and have learned how to mimic humans.

	A meticulous study of the rise of social bots is presented in \cite{ferrara2016rise}. Social bots are designed to interact with other users \citep{hwang2012socialbots, messias2013you, boshmaf2013design}, and post human-like content \citep{freitas2015reverse}. Not only do bots \emph{look} more human, but turn out to \emph{be} half-human: \cite{chu2012detecting} mentions the concept of \emph{cyborgs}, where a real person manages dozens of otherwise automated accounts. Such hybrid accounts\footnote{See the example $@TeamTrumpRussia$, https://twitter.com/teamtrumprussia.}  make the detection task extremely challenging \citep{zangerle2014sorry,ref:bbmg17}. 
	
	Bot detectors also became more sophisticated, from the earliest instances \citep{yardi2009detecting} to the current state of the art \citep{davis2016botornot} currently used in most applications today \citep{ferrara2017disinformation, monsted2017evidence, vosoughi2018spread, badawy2018analyzing}. In \cite{ferrara2016rise} the authors present a taxonomy of bot detectors, from crowd-sourcing  \citep{wang2012social, stein2011facebook, elovici2014ethical, reportSpam, botSpot} and honeypot traps \citep{lee2011seven, paradise2017creation}, to user feature oriented classifiers \citep{davis2016botornot, chu2012detecting, benevenuto2010detecting, wang2010detecting, egele2013compa, viswanath2014towards, thomas2011suspended}. Both approaches treat accounts individually, but do not detect coordinated attacks. Extant work exists for coordinated attacks, a few instances of which are \emph{CopyCatch} for Facebook 'liked' pages \citep{beutel2013copycatch}, Twitter memes \citep{ratkiewicz2011detecting}, and more generally \emph{Sybil detection} in online communities \citep{benevenuto2009detecting, aggarwal2014data, cao2014uncovering, yang2014uncovering, ghosh2012understanding, tran2009sybil, yu2008sybillimit, danezis2009sybilinfer, yu2006sybilguard, wang2013you, alvisi2013sok, cao2012aiding}.   Our  algorithm is strongly inspired from  work on image segmentation from \cite{zabih2004spatially}, which was also applied to social networks in \cite{marks2017building} to detect user location.   To the best of our knowledge, no  model for bot-human \emph{heterophilic} interactions on social networks exist, nor an efficient algorithm for joint bot detection based on such a model.

	
	\section{Data}\label{sec:data}
	
	We collected Twitter data from six different controversial events that occurred in a variety of nations (US, France, Hungary), over various time periods (2015 to 2018), and for different durations.  Some of the events were elections in the US and Hungary.  Others were for politically motivated conspiracy theories or scandals, such as Pizzagate and Macron Leaks.  Finally, there were activist groups such as Black Lives Matter (BLM).
	Details of the data are shown in Table \ref{table:dataSets}. The data was collected using Twitter's REST and STREAM APIs. For the Pizzagate, BLM 2015 and Hungary Election datasets, we collected tweets containing relevant keywords.  The US election dataset is from \cite{PDI7IN_2016}. The Macron Leaks and BLM 2016 were collected by \cite{macronLeaksDataSet,blm2016DataSet}.  In addition to the basic content information, for each tweet we also collected the user profile.

	\footnotesize
	
	\begin{table}[!hbt] \centering
		\caption{Twitter dataset descriptions.}
		\label{table:dataSets}
		\centering
		\begin{tabular}{lllc}
			\toprule
			\multicolumn{4}{c}{Descriptive statistics}                   \\
			\cmidrule(r){2-4}
			Dataset     &  Description &  Time-period  &  Number of tweets/users\\
			\midrule
			Pizzagate & Pizzagate conspiracy  & Nov-Dec 2016 &  1,025,911 / 176,822 \\
			BLM 2015 & Black Lives Matter  & Jan-Dec 2015 &  477,344 / 242,164\\
			US election  & First presidential debate & Sep-Oct 2016 & 2,435,886 / 995,918 \\
			Macron Leaks & French election scandal  & May 2017 &   570,295 / 150,848\\
			Hungary & Parliamentary election & Apr 2018 &  504,170 / 198,433\\
			BLM 2016 & Black Lives Matter & Sep 2016 &  1,274,596 / 545,937  \\
			\bottomrule
		\end{tabular}
	\end{table} 

	\normalsize
	
	\subsection{Ground Truth Construction.}
	To obtain a ground truth for bot identities, we manually labeled approximately 300 accounts per event.  The accounts were randomly selected and only required to have a minimum activity level. For each account, the labeler was given three options: human, bot, or no idea. The decision is guided by the activity patterns (number of retweets versus original tweets, volume, etc.), content of tweets (level of creativity, replies, topic diversity), and other profile features (pictures, followers/friends ratio, etc.).  Some events occurred two or three years ago, but the labeling is based on the most recent 2018 activity.  Therefore,  labelers were also given quick statistics about the account's activity during the event, but were explicitly told not to weigh it too much into their final decision.  Statistics for the number of accounts labeled as bots on each event are found in Table \ref{table:data_hand}.  We found that approximately 10\% of the accounts were labeled bots across the different datasets.
	
	\footnotesize
	\begin{table}[!hbt]
		\centering
		\caption{Statistics for hand labeled accounts for each dataset.} \label{table:data_hand}	
		\footnotesize
		\begin{tabular}{|l|c|c|c|c|c|c|}
			\toprule
			Dataset &  Pizzagate & BLM2015 & US Election & Macron Leaks & Hungary & BLM2016 \\	 \hline
			Bot labels /  &  23/304 & 21/262 &  30/300 &  19/256 &  24/300  & 30/285 \\
			Total number of accounts&&&&&&\\
			\hline
		\end{tabular}
	\end{table}
	\normalsize

	\subsection{Heterophily. }\label{subsubsec:heterophily}
\begin{figure}[!h]
		\centering
		\includegraphics[scale=0.4]{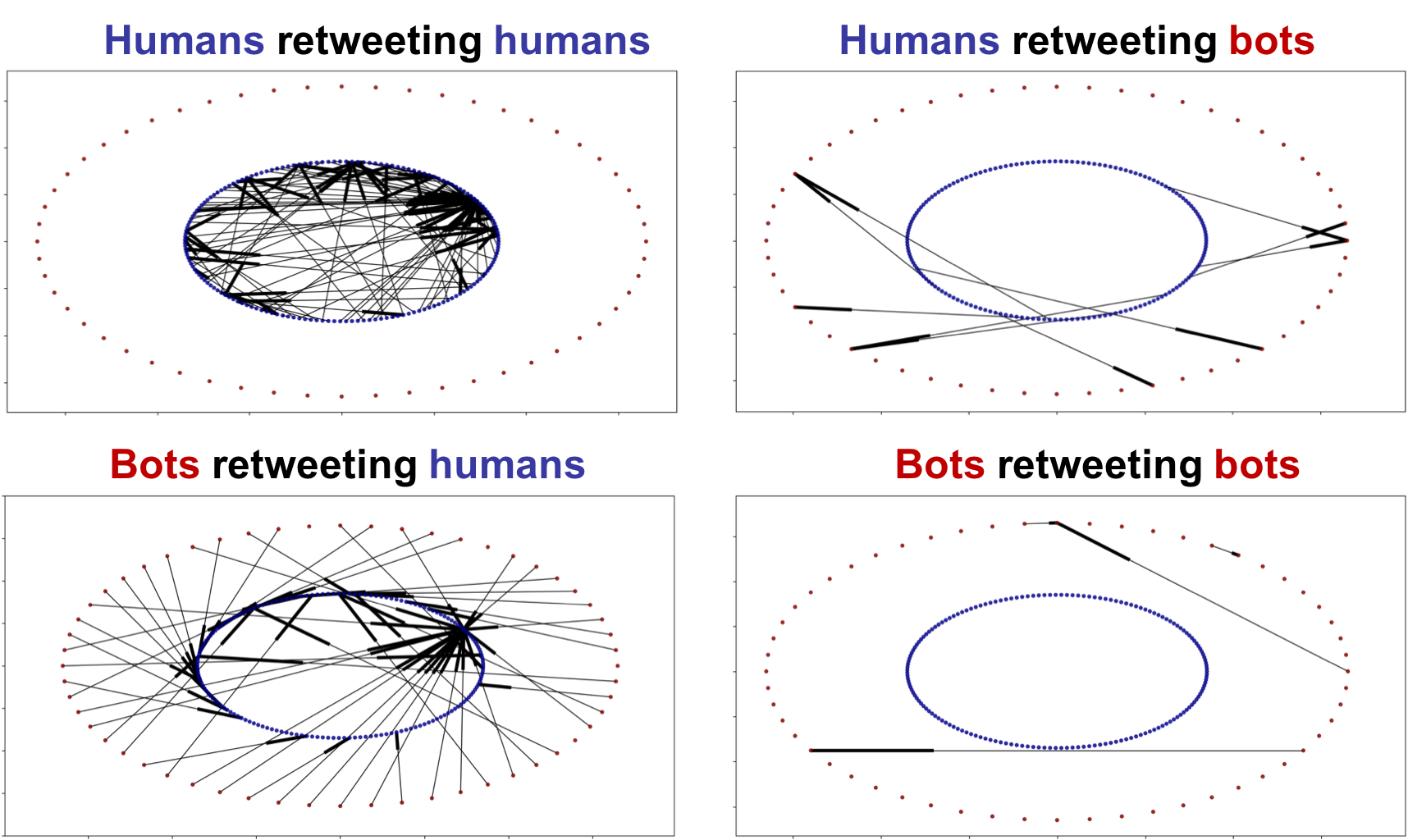} \\
		\caption{Retweet interaction networks for Pizzagate accounts labeled by humans.  Each network consists of edges between a certain pair of account types (bots and humans).} \label{fig:hetero_pizzagate}
	\end{figure}

	We study the collective behavior of the bots.  One such behavior concerns
	who the bots interact with or retweet.  Figure \ref{fig:hetero_pizzagate} shows different networks of retweeting interactions in the Pizzagate dataset  by accounts labeled by humans (see Table \ref{table:data_hand}).
	We place the bots on the peripheral ring,  humans in the central ring, and draw directed edges for the retweet interactions.  Each network corresponds to different node types on the edges.  We see qualitatively that bots prefer to retweet humans than other bots, and that humans prefer to retweet other humans instead of bots. This  phenomenon where members of a group do not interact with each other, but do interact with members of different groups is known as \emph{heterophily}.

	To obtain a more quantitative measure of heterophily,
	we broaden the set of ground truth labeled accounts by using  a popular bot detection algorithm from \cite{davis2016botornot}.  This algorithm provides a probability of being a bot.  We chose 0.5 as our threshold for being a bot. Then for each account,
	we calculated the total number of retweets it gives to humans/bots, normalized by the number of unique humans/bots it retweets.  This measures  the average number of retweets per human and retweets per bot for each account.  We refer to this quantity as the \emph{retweet rate}.
	We then look at the distribution of the retweet rates conditioned on the nodes on each end of the retweet edges: human to human, human to bot, bot to human, and bot to bot.
	In Table \ref{table:KS_hetero}  we show the mean retweet rates  for the different retweet types.  As can be seen,
	the bot to human mean is much higher than the bot to bot mean.  Also, the human to human mean is higher than the human to bot mean.  To further quantify the difference, we perform a Kolmogorov-Smirnov (KS) test on each pair of retweet rate distributions.
	The p-value for the tests on each dataset are also shown in Table \ref{table:KS_hetero}.  We find that there is a statistically significant difference between the distributions.  This supports the hypothesis that there is heterophily for the bots.  It also suggests that the humans exhibit homophily, meaning they are more likely to interact with each other than with bots.  We will use these properties to design our bot detection algorithm.
	
	\footnotesize
	\begin{table}[!hbt] \centering
		\caption{Retweet rate and p-value of KS test for different rewteet networks.  B stands for bot and H stands for human.   The zeros for the p-value mean that the p-value was less than $10^{-6}$. }
		\label{table:KS_hetero}
		\centering
		\begin{tabular}{lc@{\hspace{0.5em}}c@{\hspace{0.5em}}c@{\hspace{0.5em}}c@{\hspace{0.5em}}c@{\hspace{0.5em}}c@{\hspace{0.5em}}c@{\hspace{0.5em}}c@{\hspace{0.5em}}}
			\toprule
			\multicolumn{1}{c}{} &
			\multicolumn{2}{c}{Bots' Mean } &
			\multicolumn{2}{c}{Humans' Mean } &
			\multicolumn{4}{c}{KS-test p-value} \\
			\multicolumn{1}{c}{} &
			\multicolumn{2}{c}{ Retweet Rate} &
			\multicolumn{2}{c}{ Retweet Rate} &
			\multicolumn{4}{c}{} \\
			\cmidrule(r){2-3}
			\cmidrule(r){4-5}
			\cmidrule(r){6-9}
			Dataset     &  B $\rightarrow$ H   &  B $\rightarrow$ B  & H $\rightarrow$ H & H $\rightarrow$ B & 
			\begin{tabular}{@{}c@{}} B $\rightarrow$B vs.  \\ 
				B $\rightarrow$H  \end{tabular}  & \begin{tabular}{@{}c@{}}H$\rightarrow$H  vs.\\ H$\rightarrow$B \end{tabular} & 
			\begin{tabular}{@{}c@{}}B$\rightarrow$H  vs.\\ H$\rightarrow$H \end{tabular} & 
			\begin{tabular}{@{}c@{}}H$\rightarrow$B vs. \\ B$\rightarrow$B \end{tabular} \\
			\midrule
			US Pizzagate &  0.99 & 0.18 &  0.98 & 0.12 &  0  & 0 & 0 & 0\\
			US BLM 2015 &  0.88 & 0.17 & 0.84 &  0.05 &  0 & 0 & 0.013 & 0\\
			US election  & 0.99 & 0.66  & 0.95& 0.35 & 0 & 0 & 0 & 0\\
			Macron Leaks  &  1.26 & 0.08 & 1.13 & 0.05 &  0 & 0 & 0 & 0\\
			Hungary &  0.73 & 0.25  & 0.75 & 0.21 & 0 & 0 & 0 & 0\\
			US BLM 2016 &  0.4& 0.23 &  0.45 & 0.15 & 0 & 0  & 0 & 0\\
			\bottomrule
		\end{tabular}
	\end{table} 
	
	\normalsize


	\section{Joint Detection Using the  Ising Model}\label{sec:ntwkModel}
	In this Section, we present our joint classification model.  We assume we have
	an interaction graph $G=(V,E)$ of accounts with vertex set $V$ (the accounts) and edge set $E$ (the interactions).  
	For each  node $u_i\in V$ we observe features $\vx_{i}$ and for each pair of users $u_i,u_j\in V$ we observe interaction features $\vz_{ij}$.  The interaction features could be the number of retweets, out degree, and  in-degree.   Each node $i$ in the network has a latent variable $\Delta_i$ which is one if $i$ is a bot and zero otherwise.  We use a factor graph to model the joint distribution of the observed features and latent variables.  We define functions $\phi(\vx_{i},\Delta_{i})$ for each $i\in V$ and  $\psi(\vz_{ij},\Delta_{i},\Delta_{j})$ for $i,j\in V$.
	We refer to $\phi$ and $\psi$ respectively as the \emph{node energy} and  \emph{link energy} functions.  Then given $N$ accounts in an interaction graph/network with node features $\mathbf X =\curly{\mathbf x_i}_{i=1}^N$ and interaction features $\mathbf Z =\curly{\mathbf z_{ij}}_{i,j=1}^N$, the probability of the observed features and latent variables is
	
	\begin{equation}
	\Pr(\mathbf{X},\mathbf{Z},\mathbf{\Delta}) = \frac{\prod_{i=1}^{N}e^{-\phi(\vx_{i},\Delta_{i})}\prod_{\{i,j: \; i < j\}}e^{-\psi(\vz_{ij},\Delta_{i},\Delta_{j})}}{\mathcal{Z}}, \label{eq: joint_dist}
	\end{equation}
	where $\mathcal{Z}$ is the partition function. A probability model of this form
	which involves products of functions of single nodes and pairs of nodes has the structure of the Ising model from statistical physics \citep{ising1925beitrag}.   
	Inference of the latent variables in general can be difficult because of the presence of the partition function.  However, it has been shown that  inference problem is much easier if one specifies certain characteristics of the model \citep{zabih2004spatially, marks2017building}, which we do next.
	
	\subsection{Model Characteristics.} \label{sec:isingModel}
	\textbf{Link Energy.}
	We begin by defining the link energy functions.  First there is the case where there is no edge between nodes $u_i$ and $u_j$.  In this case we assume that we can infer nothing about the latent variables $\Delta_i$ and $\Delta_j$, so we set the link energy to be independent of the latent variables.  For simplicity, we assume that $\psi(\mathbf z_{ij},\Delta_i,\Delta_j)=0$ when there is no edge between $i$ and $j$.
	
	For pairs of nodes with an edge, we choose a functional form for the link energy.  Suppose node $u_1$ has $z_1$ retweets (its out-degree),  node $u_2$ receives  $z_2$ retweets (its in-degree), and $u_1$ retweets $u_2$ a total of $w_{12}$ times.
	If either degree is small, then the retweet edge from $u_1$ to $u_2$ provides little information about node labels because occasional retweets are very common on Twitter. For example, let $u_1$ be a  human that retweeted  $u_2$ once. If $u_1$ retweeted no one else, the edge $(u_1,u_2) $ would increase the probability that it is a bot, though it is not. In simpler terms, the only edges that contain information are the ones where $u_2$ happens to be a popular target and  $u_1$ a suspiciously active retweeter.
	Using these insights and following \cite{marks2017building} we define the link energy as
	\[
	\psi(\vz_{1,2},\Delta_1,\Delta_2|u_1 \rightarrow u_2) = \frac{w_{12}\gamma}{1+\exp((\alpha_{1}/z_1-1)+(\alpha_{2}/z_{2}-1))},
	\]
	where  $\gamma$ controls the weight of the link energy relative to the node energy, and  $\alpha_{1}, \alpha_{2}$ represent thresholds for the in  and out-degrees.  
	
	As mentioned in Section \ref{subsubsec:heterophily}, we require heterophily in our model between the bots, and homophily between the humans.  This imposes the following constraint on the link energies when there is an edge from $u_i$ to $u_j$ (a high energy corresponds to small likelihood):
	
	\begin{equation}\label{item:heterophily}
	\psi(\vz_{ij},1,0 | u_{i} \rightarrow u_{j})  \leq \psi(\vz_{ij},0,0 |   u_{i} \rightarrow u_{j}) \leq \psi(\vz_{ij},1,1 | u_{i} \rightarrow u_{j})  \leq \psi(\vz_{ij},0,1 |  u_{i} \rightarrow u_{j}). 
	\end{equation}
	The constraints above simply say that if $u_i$ retweets $u_j$,  the most likely labeling is $u_i$ is a bot and $u_j$ is a human ($\psi(\vz_{ij},1,0 | u_{i} \rightarrow u_{j}) $), and the least likely is $u_i$ is a human and $u_j$ is a bot ($\psi(\vz_{ij},0,1 | u_{i} \rightarrow u_{j})$ ).  To parameterize these energies, we introduce the constants $\epsilon < \lambda_2 < \lambda_1 \in [0,1] $ and set the link energies to
	$$
	\psi_{ij} =  \frac{w_{ij}\gamma}{1+\exp((\alpha_{1}/z_{i}-1)+(\alpha_{2}/z_{j}-1))}, \quad
	\left\{
	\begin{array}{ll}
	\psi(\vz_{ij},0,1 | u_{i} \rightarrow u_{j})   &=  \psi_{ij} \\
	\psi(\vz_{ij},1,1 | u_{i} \rightarrow u_{j})   &=  \lambda_1\psi_{ij} \\
	\psi(\vz_{ij},0,0 | u_{i} \rightarrow u_{j})   &=  \lambda_2\psi_{ij} \\
	\psi(\vz_{ij},1,0 | u_{i} \rightarrow u_{j})   &=  \epsilon\psi_{ij}.
	\end{array}
	\right.
	$$\label{alambda}
	
	For pairs of nodes with reciprocate edges $u_i \rightarrow u_j$ and $u_j \rightarrow u_i$, the energy of the pair is the aggregate of the two one-way energies. In order to efficiently infer the latent labels, the constants
	$\lambda_1, \lambda_2$ and $\epsilon $ must satisfy a set of constraints  that we present in Section \ref{sec:algo}.
	
	\paragraph{Node Energy.}  We choose a relatively simple form for the node energy.
	If we have no prior knowledge  on which accounts are bots, we set $ \phi(\vx_{i},1) = \phi(\vx_{i},0)=0$.  If we have prior knowledge, perhaps from an expert or another bot detection algorithm, we can use it for the node energy.
	However, it turns out the simple uninformative energy performs quite well.


	\section{Inference via Minimum Cut}\label{sec:algo}
	To find the most likely values for the labels given observed values $\mathbf{X}$ and $\mathbf{Z}$,  we want to maximize equation \eqref{eq: joint_dist}, which is equivalent to solving
	\begin{equation}
	\minimize_{\mathbf{\Delta}}  \sum_{i}\phi(\vx_{i},\Delta_{i}) + \sum_{i < j}\psi(\vz_{ij},\Delta_{i},\Delta_{j}). \label{eq: energy_optimization}
	\end{equation}\label{item:optim_equation}

	It has been shown that performing the optimization in equation \eqref{eq: energy_optimization} is equivalent to finding the minimum capacity cut on an appropriately defined graph \citep{boykov2001fast}.   The expression in equation \eqref{eq: energy_optimization} can be viewed as the \emph{energy } of the label configuration,  so the graph on which the minimum cut is equal to the maximum likelihood label set is referred to as the \emph{energy graph}.  We now show how to construct this graph.
	
\begin{figure}[!h]
		\centering
		\includegraphics[scale=0.65]{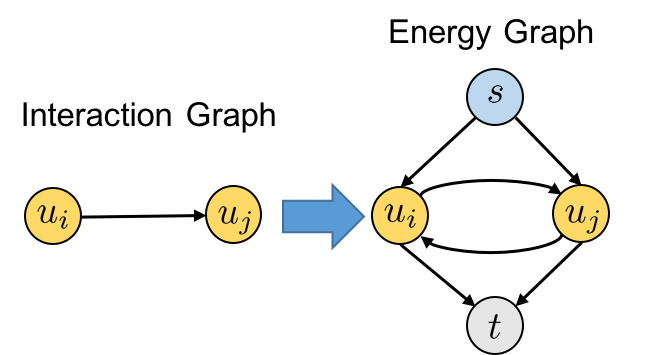} 
		\caption{Energy graph representation of the label configuration energy.} \label{fig:min-cut-graph}
	\end{figure}

	Figure \ref{fig:min-cut-graph} illustrates how to map an interaction network/graph into an energy graph.  There is a source node $s$ and a sink node $t$ in the energy graph.  There are three types of edges in the energy graph.  For each node $u_i$ in the interaction graph, there is an edge from the source and an edge to the sink: $(s,u_i)$ and $(u_i,t)$.  There are also edges between every pair of nodes with an interaction edge. For each node $u_i$ in the interaction graph,  every valid $s-t$ cut in the energy graph must either cut the edge $(s,u_i)$ or $(u_i,t)$.  If $(s,u_i)$ is cut, then $\Delta_i = 0$ and $u_i$ is a human.
	Otherwise, $(u_i,t)$ is cut, $\Delta_i=1$ and $u_i$ is a bot.  This is how a cut in the energy graph maps to a label configuration.  By proper choice of the edge weights, a minimum weight cut in the energy graph will correspond to
	a maximum likelihood label configuration.  We now define the edge
	weights following the results in \cite{boykov2001fast}.

	For each pair of users $u_i,u_j$  with an edge from $u_i$ to $u_j$ in the interaction graph, we add edges $(u_i,u_j)$ and $(u_j,u_i)$ in the energy graph with weights:
	\[
	c_{(u_{i},u_{j})}  = c_{(u_{j},u_{i})} = \frac{1}{2}\left(\psi(\vz_{ij},1,0) + \psi(\vz_{ij},0,1) - \psi(\vz_{ij},0,0) - \psi(\vz_{ij},1,1) \right).
	\]
	For each user $u_i$, the weight of the edge $(s,u_i)$ is
	$c_{(s,u_{i})}= c_{(s,u_{i})}^{out}+c_{(s,u_{i})}^{in}$, 	where 
	\[
	c_{(s,u_{i})}^{out} = \frac{1}{2}\phi(\vx_{i},0)+\frac{1}{2}\sum_{\{j:i\rightarrow j\}}\psi(\vz_{ij},0,0)+\frac{1}{4}\sum_{\{j:i\rightarrow j\}}(\psi(\vz_{ij},0,1)-\psi(\vz_{ij},1,0)),	
	\]
	and 
	\[
	c_{(s,u_{i})}^{in} = \frac{1}{2}\phi(\vx_{i},0)+\frac{1}{2}\sum_{\{j:i\leftarrow j\}}\psi(\vz_{ij},0,0)+\frac{1}{4}\sum_{\{j:i\leftarrow j\}}(\psi(\vz_{ij},0,1)-\psi(\vz_{ij},1,0)).
	\] 
	For each user $u_i$, the weight of the edge $(u_i,t)$ is
	$c_{(u_{i},t)} = c_{(u_{i},t)}^{out}+c_{(u_{i},t)}^{in}$,
	where
	\[
	c_{(u_{i},t)}^{out} = \frac{1}{2}\phi(\vx_{i},1)+\frac{1}{2}\sum_{\{j:i\rightarrow j\}}\psi(\vz_{ij},1,1)+\frac{1}{4}\sum_{\{j:i\rightarrow j\}}(\psi(\vz_{ij},1,0)-\psi(\vz_{ij},0,1)),
	\] 
	and 
	\[
	c_{(u_{i},t)}^{in} = \frac{1}{2}\phi(\vx_{i},1) +\frac{1}{2}\sum_{\{j:i\leftarrow j\}}\psi(\vz_{ij},1,1)+\frac{1}{4}\sum_{\{j:i\leftarrow j\}}(\psi(\vz_{ij},1,0)-\psi(\vz_{ij},0,1)).
	\]
	It can easily be shown that with these edge weights, the weight of an $s-t$ cut
	in the energy graph equals the energy of the corresponding label configuration in the interaction graph.
	
	\textbf{Energy Constraints.} In order for the minimum weight cut to be the most likely labels, we need the energy graph edge weights to be positive.  Also, from \cite{boykov2001fast}, it was shown that the link energies had to satisfy the condition $\psi(\vz,1,1) + \psi(\vz,0,0) \leq \psi(\vz,1,0) + \psi(\vz,0,1)$.  This condition made the energy function from equation \eqref{eq: energy_optimization} submodular and also solvable via minimum cut.  Finally, we have the heterophily constraint from equation \eqref{item:heterophily}.
	We express these constraints in terms of $\lambda_1, \lambda_2$ ,and $ \epsilon$.  First, heterophily gives  
	\begin{equation}
	0< \epsilon < \lambda_2 < \lambda_1 < 1   \label{item:ineq_heterophily}.
	\end{equation}
	Submodularity gives 
	\begin{equation}
	\epsilon =  \lambda_2 + \lambda_1 - 1 + \delta  \label{item:ineq_submodularity},
	\end{equation}
	for some $\delta \geq 0 $.  We choose $\delta = 0$ and find that edge weight positivity gives
	\begin{equation}
	0 \leq \lambda_2 - \frac{1}{2}(1- \epsilon)  \Longleftrightarrow 2 \leq 3\lambda_2 + \lambda_1. \label{item:ineq_edgePos}
	\end{equation}

	\section{Results }\label{sec:results}
	In this section, we present the results of the performance of our bot detection algorithm based on the human labelers' ground truth. In addition to the 300 hand labeled accounts for each event, we also include all the Twitter verified users. Verified users have undergone robust identity checks from Twitter in order to receive such label. This increases the number of ground-truth labels by approximately one to two thousand. We compare our performance to the state of the art bot detection algorithm known as BotOrNot \citep{davis2016botornot}. In the results presented, $(\alpha_1, \alpha_2)=(100,100)$, except for BLM2016, where the volume of interactions is much bigger.  There we set $(\alpha_1,\alpha_2)=(100,1000)$. For all events we set $\gamma=1$, $\lambda_1 = 0.8$, $\lambda_2=0.6$ and  $\epsilon=0.4$. $(\lambda_1,\lambda_2)$ is picked as the barycenter of the feasible region defined by constraints \eqref{item:ineq_heterophily}, \eqref{item:ineq_edgePos} while $\epsilon$ is determined by setting $\delta = 0$ in \eqref{item:ineq_submodularity} as we want $\epsilon$ as small as possible. As we consider the case of null node energies $\phi(\vx_{i},\Delta_{i})=0$, the choice of $\gamma$ does not impact the optimal cut, and only intervenes in computing the marginal probability of belonging to each class. We do not tune this parameter, and fix it to $\gamma = 1$, in order for the link energy differences to scale with the number of retweets $w_{ij}$.

	\begin{figure}[!h]
		\centering
		\includegraphics[scale=0.4]{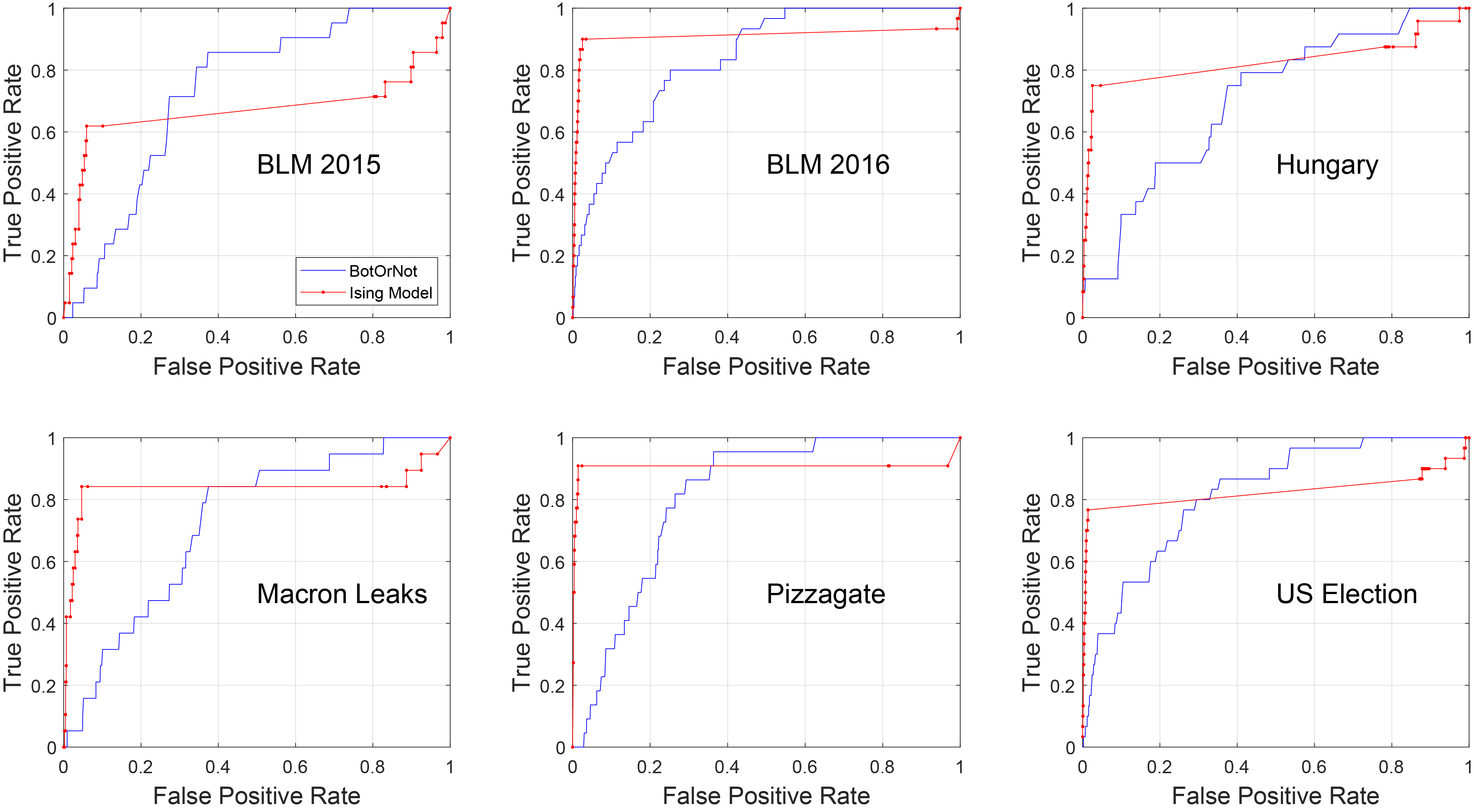} 
		\caption{ROC curves for bot detection algorithms on different Twitter datasets.} \label{fig:ROCs}
	\end{figure}

	\begin{table}[!hbt]
		\centering
		\footnotesize
		\caption{AUC and runtime for our Ising model based bot detection algorithm and BotOrNot on different Twitter datasets.} \label{table:runtimeVSaccuracy}
		\begin{tabular}{l cc lc}
			\toprule
			&
			\multicolumn{2}{c}{AUC} &
			\multicolumn{2}{c}{Runtime} \\
			\cmidrule(r){2-3}
			\cmidrule(r){4-5}
			Dataset &  \begin{tabular}{@{}c@{}}Ising \\ Model\end{tabular} & \begin{tabular}{@{}c@{}}Bot \\ or Not\end{tabular}  
			&
			\begin{tabular}{@{}c@{}}Ising \\ Model\end{tabular} & 	\begin{tabular}{@{}c@{}}Bot \\ or Not\end{tabular} \\ \hline
			Pizzagate &  \textbf{0.91}  &  0.81 & \textbf{$\approx$5 min} & 3.5s$\times$170,000    \\ 
			BLM 2015   &   0.67  &    \textbf{0.73}  & \textbf{$\approx$10 min} & 3.5s$\times$240,000  \\ 
			US election &  \textbf{0.83} &  0.82 & \textbf{$\approx$20 min} & 3.5s$\times$300,000 \\ 
			Macron Leaks    &  \textbf{0.84}  &  0.72  & \textbf{$\approx$5 min} & 3.5s$\times$150,000   \\ 
			Hungary    &  \textbf{0.83} & 0.71 & \textbf{$\approx$5 min} & 3.5s$\times$200,000  \\ 
			BLM 2016    &   \textbf{0.91}  &      0.84    & \textbf{$\approx$4 hours}  & 3.5s$\times$545,000   
		\end{tabular}
	\end{table}
	To evaluate the algorithms performance, we use the inferred labels to calculate the marginal probability of being a bot for each node.  This allows us to calculate receiver operating characteristic (ROC) curves for our algorithm and compare it to BotOrNot.  We show the ROC curves in Figure \ref{fig:ROCs}.  In
	Table \ref{table:runtimeVSaccuracy} we present the area under the curve (AUC) metric for the algorithms along with their run-time.  We note that for BotOrNot, the run-time is the time it takes to query its API, which may include additional delays beyond computation time.  Therefore, the runtimes for BotOrNot are an upper bound to the actual computational time.  From Figure \ref{fig:ROCs} and Table \ref{table:runtimeVSaccuracy} it can be seen that our algorithm usually performs better than BotOrNot in terms of AUC.  The one exception is BLM 2015.  We suspect here that the coordinated bot campaigns we are trying to detect were not as prevalent.  Our algorithm runs on several hundred thousand accounts in several minutes to a few hours.  
	
	We find that we have much fewer false positives than BotOrNot.  One of the reason is that BotOrNot labels news broadcasting accounts as bots.  Examples of this in our datasets include the Twitter accounts for Newsweek, CNN, and the New York Times.\footnote{As per the documentation on their website, bot scores were updated on May 10, 2018. The later problem seems to have been fixed, since the aforementioned verified news broadcasting accounts are now given low bot scores. However, regular users that have no bad intentions, but use automation softwares to gain popularity would still be mistaken with bots, although they are in no way involved in any influence campaign.} Because our Ising model algorithm looks for who the accounts interact with and also who interacts with the accounts, we do not mistake these trusted news sources for bots.  BotOrNot uses mainly the account behavior, so the high activity of these accounts make them appear like bots.

	In addition to the increased accuracy and fast runtime, our algorithm relies on much less information than BotOrNot (recall that we use uninformative node energies).  This shows that there is a great deal of information in the interaction graph which maps out the retweet behavior of the accounts.
	
\begin{figure}[!hbt] \centering
		\centering
		\begin{tabular}{cc}
			\includegraphics[width=0.42\textwidth]{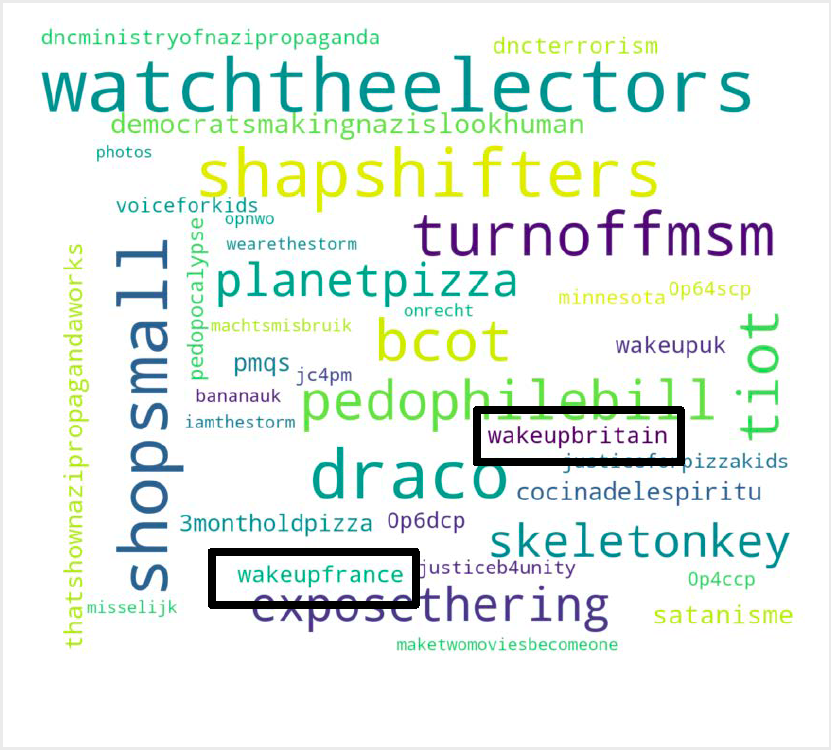} & \includegraphics[width=0.45\textwidth]{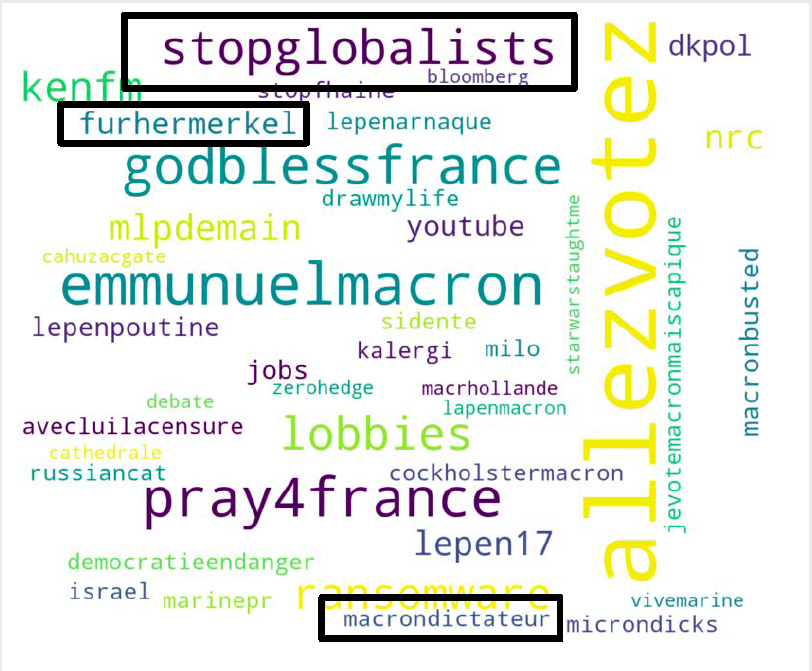} \\
		\end{tabular}
		\caption{Word clouds of hashtags used during (left) Pizzagate and (right) Macron Leaks by bots and not by humans.}\label{figure:wordClouds}
	\end{figure}

	\textbf{Bot Agenda.}  A way to understand the underlying agenda of bot designers is by looking at the content of the tweets posted by the bots. After applying our bot detection algorithm to an event, we look at the hashtags used by the bots, but not by the humans.  Figure \ref{figure:wordClouds} illustrates the results of such analysis on the Pizzagate and Macron Leaks datasets. 
	
	In Pizzagate, the bots try to export the controversy outside of the US, by transforming the usual \emph{$\#$WakeUpAmerica} into \emph{\# WakeUpBritain}, and \emph{$\#$WakeUpFrance}. Similar intent can be seen in the Macron Leaks analysis, with the use of the \emph{$\#$furherMerkel} and \emph{$\#$stopGlobalists} hashtags.  In both cases, it looks like bots intend to internationalize an otherwise local controversy in order to reach a broader audience, especially that of western Europe. It appears as if the bots are trying to shift the theme of the conversation to certain types of controversies (instability of Europe, Brexit, Frexit, etc.).

	\section{Conclusion}
		
	We have presented an Ising model for the probability of bot labels
	in a social network.  Our model was motivated by the observation
	of heterophily in bot interactions.  We infer the bot labels
	efficiently by solving a minimum cut problem.  We find our joint detection
	algorithm outperforms current existing algorithms.  This may be due
	to the fact that the bots are coordinated. Individual detection does not
	allow one to take advantage of the coordinated behavior.  We also find that our algorithm is able to discern bot accounts from just active trusted news source accounts, in contrast to other popular algorithms.  This is a result of our incorporation of the joint interactions of all accounts.    Analysis of the content of the bots reveals that they seem to have an agenda focused on controversial topics.  We feel that for these types of coordinated campaigns, joint detection is much more effective than individual detection.
	
	\subsubsection*{Acknowledgments}
	This research was supported in part by the Office of Naval Research (ONR) and Charles Stark Draper Laboratory, Inc (Draper). The views presented here are those of the author and do not necessarily represent those of ONR, Draper, or MIT.

	\newpage
	\small
	\bibliography{bot}

\begin{thebibliography}{67}
\providecommand{\natexlab}[1]{#1}
\providecommand{\url}[1]{\texttt{#1}}
\expandafter\ifx\csname urlstyle\endcsname\relax
  \providecommand{\doi}[1]{doi: #1}\else
  \providecommand{\doi}{doi: \begingroup \urlstyle{rm}\Url}\fi

\bibitem[Aggarwal(2014)]{aggarwal2014data}
Charu~C Aggarwal.
\newblock \emph{Data classification: algorithms and applications}.
\newblock CRC Press, 2014.

\bibitem[Alvisi et~al.(2013)Alvisi, Clement, Epasto, Lattanzi, and
  Panconesi]{alvisi2013sok}
Lorenzo Alvisi, Allen Clement, Alessandro Epasto, Silvio Lattanzi, and
  Alessandro Panconesi.
\newblock Sok: The evolution of sybil defense via social networks.
\newblock In \emph{Security and Privacy (SP), 2013 IEEE Symposium on}, pages
  382--396. IEEE, 2013.

\bibitem[Badawy et~al.(2018)Badawy, Ferrara, and Lerman]{badawy2018analyzing}
Adam Badawy, Emilio Ferrara, and Kristina Lerman.
\newblock Analyzing the digital traces of political manipulation: The 2016
  russian interference twitter campaign.
\newblock \emph{arXiv preprint arXiv:1802.04291}, 2018.

\bibitem[Benevenuto et~al.(2009)Benevenuto, Rodrigues, Almeida, Almeida, and
  Gon{\c{c}}alves]{benevenuto2009detecting}
Fabr{\'\i}cio Benevenuto, Tiago Rodrigues, Virg{\'\i}lio Almeida, Jussara
  Almeida, and Marcos Gon{\c{c}}alves.
\newblock Detecting spammers and content promoters in online video social
  networks.
\newblock In \emph{Proceedings of the 32nd international ACM SIGIR conference
  on Research and development in information retrieval}, pages 620--627. ACM,
  2009.

\bibitem[Benevenuto et~al.(2010)Benevenuto, Magno, Rodrigues, and
  Almeida]{benevenuto2010detecting}
Fabricio Benevenuto, Gabriel Magno, Tiago Rodrigues, and Virgilio Almeida.
\newblock Detecting spammers on twitter.
\newblock In \emph{Collaboration, electronic messaging, anti-abuse and spam
  conference (CEAS)}, volume~6, page~12, 2010.

\bibitem[Beutel et~al.(2013)Beutel, Xu, Guruswami, Palow, and
  Faloutsos]{beutel2013copycatch}
Alex Beutel, Wanhong Xu, Venkatesan Guruswami, Christopher Palow, and Christos
  Faloutsos.
\newblock Copycatch: stopping group attacks by spotting lockstep behavior in
  social networks.
\newblock In \emph{Proceedings of the 22nd international conference on World
  Wide Web}, pages 119--130. ACM, 2013.

\bibitem[Boshmaf et~al.(2013)Boshmaf, Muslukhov, Beznosov, and
  Ripeanu]{boshmaf2013design}
Yazan Boshmaf, Ildar Muslukhov, Konstantin Beznosov, and Matei Ripeanu.
\newblock Design and analysis of a social botnet.
\newblock \emph{Computer Networks}, 57\penalty0 (2):\penalty0 556--578, 2013.

\bibitem[Boykov et~al.(2001)Boykov, Veksler, and Zabih]{boykov2001fast}
Yuri Boykov, Olga Veksler, and Ramin Zabih.
\newblock Fast approximate energy minimization via graph cuts.
\newblock \emph{IEEE Transactions on pattern analysis and machine
  intelligence}, 23\penalty0 (11):\penalty0 1222--1239, 2001.

\bibitem[Byrnes(2016)]{byrnes2016bot}
Nanette Byrnes.
\newblock How the bot-y politic influenced this election.
\newblock \emph{Technology Rev.}, 2016.

\bibitem[Cao et~al.(2012)Cao, Sirivianos, Yang, and Pregueiro]{cao2012aiding}
Qiang Cao, Michael Sirivianos, Xiaowei Yang, and Tiago Pregueiro.
\newblock Aiding the detection of fake accounts in large scale social online
  services.
\newblock In \emph{Proceedings of the 9th USENIX conference on Networked
  Systems Design and Implementation}, pages 15--15. USENIX Association, 2012.

\bibitem[Cao et~al.(2014)Cao, Yang, Yu, and Palow]{cao2014uncovering}
Qiang Cao, Xiaowei Yang, Jieqi Yu, and Christopher Palow.
\newblock Uncovering large groups of active malicious accounts in online social
  networks.
\newblock In \emph{Proceedings of the 2014 ACM SIGSAC Conference on Computer
  and Communications Security}, pages 477--488. ACM, 2014.

\bibitem[Castillo et~al.(2007)Castillo, Donato, Gionis, Murdock, and
  Silvestri]{castillo2007know}
Carlos Castillo, Debora Donato, Aristides Gionis, Vanessa Murdock, and Fabrizio
  Silvestri.
\newblock Know your neighbors: Web spam detection using the web topology.
\newblock In \emph{Proceedings of the 30th annual international ACM SIGIR
  conference on Research and development in information retrieval}, pages
  423--430. ACM, 2007.

\bibitem[Chu et~al.(2012)Chu, Gianvecchio, Wang, and Jajodia]{chu2012detecting}
Zi~Chu, Steven Gianvecchio, Haining Wang, and Sushil Jajodia.
\newblock Detecting automation of twitter accounts: Are you a human, bot, or
  cyborg?
\newblock \emph{IEEE Transactions on Dependable and Secure Computing},
  9\penalty0 (6):\penalty0 811--824, 2012.

\bibitem[Danezis and Mittal(2009)]{danezis2009sybilinfer}
George Danezis and Prateek Mittal.
\newblock Sybilinfer: Detecting sybil nodes using social networks.
\newblock In \emph{NDSS}, pages 1--15. San Diego, CA, 2009.

\bibitem[Davis et~al.(2016)Davis, Varol, Ferrara, Flammini, and
  Menczer]{davis2016botornot}
Clayton~Allen Davis, Onur Varol, Emilio Ferrara, Alessandro Flammini, and
  Filippo Menczer.
\newblock Botornot: A system to evaluate social bots.
\newblock In \emph{Proceedings of the 25th International Conference Companion
  on World Wide Web}, pages 273--274. International World Wide Web Conferences
  Steering Committee, 2016.

\bibitem[Egele et~al.(2013)Egele, Stringhini, Kruegel, and
  Vigna]{egele2013compa}
Manuel Egele, Gianluca Stringhini, Christopher Kruegel, and Giovanni Vigna.
\newblock Compa: Detecting compromised accounts on social networks.
\newblock In \emph{NDSS}, 2013.

\bibitem[Elovici et~al.(2014)Elovici, Fire, Herzberg, and
  Shulman]{elovici2014ethical}
Yuval Elovici, Michael Fire, Amir Herzberg, and Haya Shulman.
\newblock Ethical considerations when employing fake identities in online
  social networks for research.
\newblock \emph{Science and engineering ethics}, 20\penalty0 (4):\penalty0
  1027--1043, 2014.

\bibitem[Fandos and Shane(2017)]{ref:russianbots_govtresponse}
Nocholas Fandos and Scott Shane.
\newblock {Senator Berates Twitter Over ‘Inadequate’ Inquiry Into Russian
  Meddling }.
\newblock \emph{The New York Times}, September 2017.
\newblock URL
  \url{https://www.nytimes.com/2017/09/28/us/politics/twitter-russia-interference-2016-election-investigation.html?mtrref=www.google.com}.

\bibitem[Ferrara(2017)]{ferrara2017disinformation}
Emilio Ferrara.
\newblock Disinformation and social bot operations in the run up to the 2017
  french presidential election.
\newblock 2017.

\bibitem[Ferrara et~al.(2016)Ferrara, Varol, Davis, Menczer, and
  Flammini]{ferrara2016rise}
Emilio Ferrara, Onur Varol, Clayton Davis, Filippo Menczer, and Alessandro
  Flammini.
\newblock The rise of social bots.
\newblock \emph{Communications of the ACM}, 59\penalty0 (7):\penalty0 96--104,
  2016.

\bibitem[Freitas et~al.(2015)Freitas, Benevenuto, Ghosh, and
  Veloso]{freitas2015reverse}
Carlos Freitas, Fabricio Benevenuto, Saptarshi Ghosh, and Adriano Veloso.
\newblock Reverse engineering socialbot infiltration strategies in twitter.
\newblock In \emph{Advances in Social Networks Analysis and Mining (ASONAM),
  2015 IEEE/ACM International Conference on}, pages 25--32. IEEE, 2015.

\bibitem[Ghosh et~al.(2011)Ghosh, Surachawala, and Lerman]{ghosh2011entropy}
Rumi Ghosh, Tawan Surachawala, and Kristina Lerman.
\newblock Entropy-based classification of'retweeting'activity on twitter.
\newblock \emph{arXiv preprint arXiv:1106.0346}, 2011.

\bibitem[Ghosh et~al.(2012)Ghosh, Viswanath, Kooti, Sharma, Korlam, Benevenuto,
  Ganguly, and Gummadi]{ghosh2012understanding}
Saptarshi Ghosh, Bimal Viswanath, Farshad Kooti, Naveen~Kumar Sharma, Gautam
  Korlam, Fabricio Benevenuto, Niloy Ganguly, and Krishna~Phani Gummadi.
\newblock Understanding and combating link farming in the twitter social
  network.
\newblock In \emph{Proceedings of the 21st international conference on World
  Wide Web}, pages 61--70. ACM, 2012.

\bibitem[Gianvecchio et~al.(2011)Gianvecchio, Xie, Wu, and
  Wang]{gianvecchio2011humans}
Steven Gianvecchio, Mengjun Xie, Zhenyu Wu, and Haining Wang.
\newblock Humans and bots in internet chat: measurement, analysis, and
  automated classification.
\newblock \emph{IEEE/ACM Transactions on Networking (TON)}, 19\penalty0
  (5):\penalty0 1557--1571, 2011.

\bibitem[Guilbeault and Woolley(2016)]{guilbeault2016twitter}
Douglas Guilbeault and Samuel Woolley.
\newblock How twitter bots are shaping the election.
\newblock \emph{The Atlantic}, 1, 2016.

\bibitem[Gy{\"o}ngyi and Garcia-Molina(2005)]{gyongyi2005web}
Zoltan Gy{\"o}ngyi and Hector Garcia-Molina.
\newblock Web spam taxonomy.
\newblock In \emph{First international workshop on adversarial information
  retrieval on the web (AIRWeb 2005)}, 2005.

\bibitem[Gy{\"o}ngyi et~al.(2004)Gy{\"o}ngyi, Garcia-Molina, and
  Pedersen]{gyongyi2004combating}
Zolt{\'a}n Gy{\"o}ngyi, Hector Garcia-Molina, and Jan Pedersen.
\newblock Combating web spam with trustrank.
\newblock In \emph{Proceedings of the Thirtieth international conference on
  Very large data bases-Volume 30}, pages 576--587. VLDB Endowment, 2004.

\bibitem[Heymann et~al.(2007)Heymann, Koutrika, and
  Garcia-Molina]{heymann2007fighting}
Paul Heymann, Georgia Koutrika, and Hector Garcia-Molina.
\newblock Fighting spam on social web sites: A survey of approaches and future
  challenges.
\newblock \emph{IEEE Internet Computing}, 11\penalty0 (6), 2007.

\bibitem[Hwang et~al.(2012)Hwang, Pearce, and Nanis]{hwang2012socialbots}
Tim Hwang, Ian Pearce, and Max Nanis.
\newblock Socialbots: Voices from the fronts.
\newblock \emph{interactions}, 19\penalty0 (2):\penalty0 38--45, 2012.

\bibitem[Ising(1925)]{ising1925beitrag}
Ernst Ising.
\newblock Beitrag zur theorie des ferromagnetismus.
\newblock \emph{Zeitschrift f{\"u}r Physik}, 31\penalty0 (1):\penalty0
  253--258, 1925.

\bibitem[Lee et~al.(2011)Lee, Eoff, and Caverlee]{lee2011seven}
Kyumin Lee, Brian~David Eoff, and James Caverlee.
\newblock Seven months with the devils: A long-term study of content polluters
  on twitter.
\newblock In \emph{ICWSM}, 2011.

\bibitem[Li et~al.(2014)Li, Mukherjee, Liu, Kornfield, and
  Emery]{li2014detecting}
Huayi Li, Arjun Mukherjee, Bing Liu, Rachel Kornfield, and Sherry Emery.
\newblock Detecting campaign promoters on twitter using markov random fields.
\newblock In \emph{Data Mining (ICDM), 2014 IEEE International Conference on},
  pages 290--299. IEEE, 2014.

\bibitem[Littman et~al.(2016)Littman, Wrubel, and Kerchner]{PDI7IN_2016}
Justin Littman, Laura Wrubel, and Daniel Kerchner.
\newblock 2016 united states presidential election tweet ids, 2016.
\newblock URL \url{https://doi.org/10.7910/DVN/PDI7IN}.

\bibitem[Malmgren et~al.(2009)Malmgren, Hofman, Amaral, and
  Watts]{malmgren2009characterizing}
R~Dean Malmgren, Jake~M Hofman, Luis~AN Amaral, and Duncan~J Watts.
\newblock Characterizing individual communication patterns.
\newblock In \emph{Proceedings of the 15th ACM SIGKDD international conference
  on Knowledge discovery and data mining}, pages 607--616. ACM, 2009.

\bibitem[Marks and Zaman(2017)]{marks2017building}
Christopher Marks and Tauhid Zaman.
\newblock Building a location-based set of social media users.
\newblock \emph{arXiv preprint arXiv:1711.01481}, 2017.

\bibitem[Messias et~al.(2013)Messias, Schmidt, Oliveira, and
  Benevenuto]{messias2013you}
Johnnatan Messias, Lucas Schmidt, Ricardo Oliveira, and Fabr{\'\i}cio
  Benevenuto.
\newblock You followed my bot! transforming robots into influential users in
  twitter.
\newblock \emph{First Monday}, 18\penalty0 (7), 2013.

\bibitem[Modaresi et~al.()Modaresi, Vielma, and Zaman]{modaresistrategic}
Sina Modaresi, Juan~Pablo Vielma, and Tauhid Zaman.
\newblock Strategic timing of content in online social networks.

\bibitem[M{\o}nsted et~al.(2017)M{\o}nsted, Sapie{\.z}y{\'n}ski, Ferrara, and
  Lehmann]{monsted2017evidence}
Bjarke M{\o}nsted, Piotr Sapie{\.z}y{\'n}ski, Emilio Ferrara, and Sune Lehmann.
\newblock Evidence of complex contagion of information in social media: An
  experiment using twitter bots.
\newblock \emph{PloS one}, 12\penalty0 (9):\penalty0 e0184148, 2017.

\bibitem[Ntoulas et~al.(2006)Ntoulas, Najork, Manasse, and
  Fetterly]{ntoulas2006detecting}
Alexandros Ntoulas, Marc Najork, Mark Manasse, and Dennis Fetterly.
\newblock Detecting spam web pages through content analysis.
\newblock In \emph{Proceedings of the 15th international conference on World
  Wide Web}, pages 83--92. ACM, 2006.

\bibitem[Paradise et~al.(2017)Paradise, Shabtai, Puzis, Elyashar, Elovici,
  Roshandel, and Peylo]{paradise2017creation}
Abigail Paradise, Asaf Shabtai, Rami Puzis, Aviad Elyashar, Yuval Elovici,
  Mehran Roshandel, and Christoph Peylo.
\newblock Creation and management of social network honeypots for detecting
  targeted cyber attacks.
\newblock \emph{IEEE Transactions on Computational Social Systems}, 4\penalty0
  (3):\penalty0 65--79, 2017.

\bibitem[Parlapiano and Lee(2018)]{ref:russianbots}
Alicia Parlapiano and C.~Lee, Jasmine.
\newblock {The Propaganda Tools Used by Russians to Influence the 2016
  Election}.
\newblock \emph{The New York Times}, February 2018.
\newblock URL
  \url{https://www.nytimes.com/interactive/2018/02/16/us/politics/russia-propaganda-election-2016.html}.

\bibitem[Price(2018)]{ref:russianbots_feinstein}
Molly Price.
\newblock Democrats urge facebook and twitter to probe russian bots.
\newblock \emph{CNET}, January 2018.
\newblock URL
  \url{https://www.cnet.com/news/facebook-and-twitter-asked-again-to-investigate-russian-bots/}.

\bibitem[Raghavan et~al.(2014)Raghavan, Ver~Steeg, Galstyan, and
  Tartakovsky]{raghavan2014modeling}
Vasanthan Raghavan, Greg Ver~Steeg, Aram Galstyan, and Alexander~G Tartakovsky.
\newblock Modeling temporal activity patterns in dynamic social networks.
\newblock \emph{IEEE Transactions on Computational Social Systems}, 1\penalty0
  (1):\penalty0 89--107, 2014.

\bibitem[Ratkiewicz et~al.(2011)Ratkiewicz, Conover, Meiss, Gon{\c{c}}alves,
  Flammini, and Menczer]{ratkiewicz2011detecting}
Jacob Ratkiewicz, Michael Conover, Mark~R Meiss, Bruno Gon{\c{c}}alves,
  Alessandro Flammini, and Filippo Menczer.
\newblock Detecting and tracking political abuse in social media.
\newblock \emph{ICWSM}, 11:\penalty0 297--304, 2011.

\bibitem[Rogers(2010)]{rogers2010diffusion}
Everett~M Rogers.
\newblock \emph{Diffusion of innovations}.
\newblock Simon and Schuster, 2010.

\bibitem[Sahami et~al.(1998)Sahami, Dumais, Heckerman, and
  Horvitz]{sahami1998bayesian}
Mehran Sahami, Susan Dumais, David Heckerman, and Eric Horvitz.
\newblock A bayesian approach to filtering junk e-mail.
\newblock In \emph{Learning for Text Categorization: Papers from the 1998
  workshop}, volume~62, pages 98--105, 1998.

\bibitem[Shane(2017)]{shane2017fake}
S~Shane.
\newblock The fake americans russia created to influence the election.
\newblock \emph{The New York Times}, 7, 2017.

\bibitem[Stein et~al.(2011)Stein, Chen, and Mangla]{stein2011facebook}
Tao Stein, Erdong Chen, and Karan Mangla.
\newblock Facebook immune system.
\newblock In \emph{Proceedings of the 4th Workshop on Social Network Systems},
  page~8. ACM, 2011.

\bibitem[Summers(2017{\natexlab{a}})]{blm2016DataSet}
Ed~Summers.
\newblock Black lives matter tweets - dataset, 2017{\natexlab{a}}.
\newblock URL
  \url{https://archive.org/details/blacklivesmatter-tweets-2016.txt}.

\bibitem[Summers(2017{\natexlab{b}})]{macronLeaksDataSet}
Ed~Summers.
\newblock Macron leaks tweets - dataset, 2017{\natexlab{b}}.
\newblock URL \url{https://archive.org/details/MacronleaksTweets}.

\bibitem[Syeed and Frier(2017)]{ref:bbmg17}
Nafeesa Syeed and Sarah Frier.
\newblock {Pro-Russian Bots Sharpen Online Attacks for 2018 U.S. Vote}.
\newblock \emph{Bloomberg}, September 2017.
\newblock URL
  \url{https://www.bloomberg.com/news/articles/2017-09-01/russia-linked-bots-hone-online-attack-plans-for-2018-u-s-vote}.

\bibitem[Thomas et~al.(2011)Thomas, Grier, Song, and
  Paxson]{thomas2011suspended}
Kurt Thomas, Chris Grier, Dawn Song, and Vern Paxson.
\newblock Suspended accounts in retrospect: an analysis of twitter spam.
\newblock In \emph{Proceedings of the 2011 ACM SIGCOMM conference on Internet
  measurement conference}, pages 243--258. ACM, 2011.

\bibitem[Tran et~al.(2009)Tran, Min, Li, and Subramanian]{tran2009sybil}
Dinh~Nguyen Tran, Bonan Min, Jinyang Li, and Lakshminarayanan Subramanian.
\newblock Sybil-resilient online content voting.
\newblock In \emph{NSDI}, volume~9, pages 15--28, 2009.

\bibitem[Twitter({\natexlab{a}})]{botSpot}
Twitter.
\newblock Twitter unofficial botspot feed, {\natexlab{a}}.
\newblock URL \url{https://twitter.com/hashtag/BotSpot}.

\bibitem[Twitter({\natexlab{b}})]{reportSpam}
Twitter.
\newblock Twitter official spam report tool, {\natexlab{b}}.
\newblock URL
  \url{https://help.twitter.com/en/safety-and-security/report-spam}.

\bibitem[Viswanath et~al.(2014)Viswanath, Bashir, Crovella, Guha, Gummadi,
  Krishnamurthy, and Mislove]{viswanath2014towards}
Bimal Viswanath, Muhammad~Ahmad Bashir, Mark Crovella, Saikat Guha, Krishna~P
  Gummadi, Balachander Krishnamurthy, and Alan Mislove.
\newblock Towards detecting anomalous user behavior in online social networks.
\newblock In \emph{USENIX Security Symposium}, pages 223--238, 2014.

\bibitem[Vosoughi et~al.(2018)Vosoughi, Roy, and Aral]{vosoughi2018spread}
Soroush Vosoughi, Deb Roy, and Sinan Aral.
\newblock The spread of true and false news online.
\newblock \emph{Science}, 359\penalty0 (6380):\penalty0 1146--1151, 2018.

\bibitem[Wang(2010)]{wang2010detecting}
Alex~Hai Wang.
\newblock Detecting spam bots in online social networking sites: a machine
  learning approach.
\newblock In \emph{IFIP Annual Conference on Data and Applications Security and
  Privacy}, pages 335--342. Springer, 2010.

\bibitem[Wang et~al.(2012)Wang, Mohanlal, Wilson, Wang, Metzger, Zheng, and
  Zhao]{wang2012social}
Gang Wang, Manish Mohanlal, Christo Wilson, Xiao Wang, Miriam Metzger, Haitao
  Zheng, and Ben~Y Zhao.
\newblock Social turing tests: Crowdsourcing sybil detection.
\newblock \emph{arXiv preprint arXiv:1205.3856}, 2012.

\bibitem[Wang et~al.(2013)Wang, Konolige, Wilson, Wang, Zheng, and
  Zhao]{wang2013you}
Gang Wang, Tristan Konolige, Christo Wilson, Xiao Wang, Haitao Zheng, and Ben~Y
  Zhao.
\newblock You are how you click: Clickstream analysis for sybil detection.
\newblock In \emph{USENIX Security Symposium}, volume~9, pages 1--008, 2013.

\bibitem[Yang et~al.(2014)Yang, Wilson, Wang, Gao, Zhao, and
  Dai]{yang2014uncovering}
Zhi Yang, Christo Wilson, Xiao Wang, Tingting Gao, Ben~Y Zhao, and Yafei Dai.
\newblock Uncovering social network sybils in the wild.
\newblock \emph{ACM Transactions on Knowledge Discovery from Data (TKDD)},
  8\penalty0 (1):\penalty0 2, 2014.

\bibitem[Yardi et~al.(2009)Yardi, Romero, Schoenebeck,
  et~al.]{yardi2009detecting}
Sarita Yardi, Daniel Romero, Grant Schoenebeck, et~al.
\newblock Detecting spam in a twitter network.
\newblock \emph{First Monday}, 15\penalty0 (1), 2009.

\bibitem[Yu et~al.(2006)Yu, Kaminsky, Gibbons, and Flaxman]{yu2006sybilguard}
Haifeng Yu, Michael Kaminsky, Phillip~B Gibbons, and Abraham Flaxman.
\newblock Sybilguard: defending against sybil attacks via social networks.
\newblock In \emph{ACM SIGCOMM Computer Communication Review}, volume~36, pages
  267--278. ACM, 2006.

\bibitem[Yu et~al.(2008)Yu, Gibbons, Kaminsky, and Xiao]{yu2008sybillimit}
Haifeng Yu, Phillip~B Gibbons, Michael Kaminsky, and Feng Xiao.
\newblock Sybillimit: A near-optimal social network defense against sybil
  attacks.
\newblock In \emph{Security and Privacy, 2008. SP 2008. IEEE Symposium on},
  pages 3--17. IEEE, 2008.

\bibitem[Zabih and Kolmogorov(2004)]{zabih2004spatially}
Ramin Zabih and Vladimir Kolmogorov.
\newblock Spatially coherent clustering using graph cuts.
\newblock In \emph{Computer Vision and Pattern Recognition, 2004. CVPR 2004.
  Proceedings of the 2004 IEEE Computer Society Conference on}, volume~2, pages
  II--II. IEEE, 2004.

\bibitem[Zangerle and Specht(2014)]{zangerle2014sorry}
Eva Zangerle and G{\"u}nther Specht.
\newblock Sorry, i was hacked: a classification of compromised twitter
  accounts.
\newblock In \emph{Proceedings of the 29th Annual ACM Symposium on Applied
  Computing}, pages 587--593. ACM, 2014.

\bibitem[Zhang and Paxson(2011)]{zhang2011detecting}
Chao~Michael Zhang and Vern Paxson.
\newblock Detecting and analyzing automated activity on twitter.
\newblock In \emph{International Conference on Passive and Active Network
  Measurement}, pages 102--111. Springer, 2011.

\end{thebibliography}
	\bibliographystyle{plainnat}

\end{document}